\documentclass[a4paper]{article}
\usepackage{Odyssey2018}
\usepackage{epsfig,amssymb,amsmath}
\usepackage{color}
\usepackage{multirow}
\usepackage[framemethod=TikZ]{mdframed}

\ninept

\mdfdefinestyle{MyFrame}{%
    linecolor=black,
    outerlinewidth=.1pt,
    roundcorner=1pt,
    innertopmargin=\baselineskip,
    innerbottommargin=\baselineskip,
    innerrightmargin=5pt,
    innerleftmargin=0pt,
    backgroundcolor=gray!10!white}

\title{Perceptual Evaluation of the Effectiveness of \\Voice Disguise by Age Modification}

\makeatletter
\def\name#1{\gdef\@name{#1\\}}
\makeatother
\name{{\em Rosa Gonz\'alez Hautam\"aki, Anssi Kanervisto, Ville Hautam\"aki, Tomi Kinnunen}\thanks{This work was partially funded by the Academy of Finland (projects no. 309629 and 313970)}}
\address{School of Computing  \\
University of Eastern Finland, Finland \\
{\small \tt \{rgonza, anssk, villeh, tkinnu\}@cs.uef.fi }}
\begin{document}
\maketitle

\begin{abstract}
\emph{Voice disguise}, the purposeful modification of one's speaker identity with the aim of avoiding being identified as oneself, is a low-effort way to fool speaker recognition, whether performed by a human or an automatic speaker verification (ASV) system. We present an evaluation of the effectiveness of \emph{age stereotypes} as a voice disguise strategy, as a follow up to our recent work where 60 native Finnish speakers attempted to sound like an \emph{elderly} and like a \emph{child}. In that study, we presented evidence that both ASV and human observers could easily miss the target speaker but we did not address how \emph{believable} the presented vocal age stereotypes were; this study serves to fill that gap.  The interesting cases would be speakers who succeed in being missed by the ASV system, and which a typical listener cannot detect as being a disguise. We carry out a perceptual test to study the quality of the disguised speech samples. The listening test was carried out both locally and with the help of Amazon's Mechanical Turk (MT) crowd-workers. A total of 91 listeners participated in the test and were instructed to estimate both the speaker's \emph{chronological} and \emph{intended} age. The results indicate that age estimations for the intended old and child voices for female speakers were towards the target age groups, while for male speakers, the age estimations corresponded to the direction of the target voice only for elderly voices. In the case of intended child's voice, listeners estimated the age of male speakers to be older than their chronological age for most of the speakers and not the intended target age.
\end{abstract}

\section{Introduction}
The human voice is highly flexible \cite{eriksson1997flexible}. Besides relaying the spoken message, the speaker can alter his or her \emph{voice quality} by changes in phonation, articulation or both \cite{eriksson2010disguised,CIVILcorpus,NTUstudy}. 
The human voice production system is not rigid and can be modified. Such modifications can be \emph{non-deliberate} or \emph{deliberate} \cite{Rodman2000computer}. The former refers to changes in conditions that are not under the speaker's conscious control (\emph{e.g.} speaker's health)    
or dependent of the environment (\emph{e.g.} Lombard reflex under noisy environments), whereas deliberate modification is actively enforced by the speaker so that he or she is fully aware of it. 
Voice acting, disguise and impersonation \cite{Leemann201597,zhang2012acoustic,hautamakiage} are good examples of this.

Speech modifications are of concern in \emph{speaker recognition} \cite{KinnunenOverview}, the task of recognizing persons from their voices. With the proliferation of mobile devices, the demand for speech technology applications has increased towards user authentication from a remote terminal.
Another application relates to law enforcement and forensics, where a speaker's voice could be used for surveillance or be subjected to forensic voice profiling. 
Whether performed by an automatic system or a listener, reliability of speaker recognition under deliberate voice modification is of great concern. From the perspective of the perpetrator, the purpose of deliberate voice modification relates to the aim of concealing one's identity. This process, known as \emph{voice disguise}, is the focus of our study.

Speakers may employ a number of strategies to disguise their voices, including the use of external objects  (mask, helmet, handkerchief, hand over mouth, pencil, chewing gum); forced modifications of the vocal cavities (pulled cheeks, pinched nose); changes in phonation (creaky, hoar, whisper); adopting a foreign language or dialect. In our recent work, we addressed the impact of voice disguise to automatic speaker verification (ASV) systems \cite{hautamakiage} and  analyzed the variations in acoustic parameters and listeners' performance \cite{gonzalezhautamaki2017}. The data used in those studies focus on the voice disguise strategy of modifying one's voice to sound like an elderly and a child. Our prior studies found the child mimicry to be particularly detrimental to the performance of ASV systems (see selected results in Table~\ref{tab:EER}) \cite{hautamakiage}, possibly due to a substantial increase in the fundamental frequency that causes a large mismatch in the mel-cepstral features employed by the ASV system. Similar results were found from perceptual experiments, where our listener panel had difficulty connecting modal and disguised samples from the same speaker.
\begin{table}[hbp]
\caption{{\it Performance of i-vector PLDA speaker recognition system, in terms of equal error rate (EER, \%), on the dataset with male and female speakers speaking in their modal and disguised voices (elderly and child) \cite{gonzalezhautamaki2017}. Additional results are available in \cite{hautamakiage}.}}
\centering
%\resizebox{\linewidth}{!}{
\begin{tabular}{|c|c|c|c|}
\cline{2-4}
\multicolumn{ 1}{l}{}& \multicolumn{ 1}{|c|}{\textbf{Modal}} & \multicolumn{1}{c|}{\textbf{Disguise}} & \multicolumn{1}{c|}{\textbf{Disguise}}  \\
\multicolumn{ 1}{l}{} & \multicolumn{ 1}{|c|}{} & \multicolumn{1}{c|}{\textbf{elderly}} & \multicolumn{1}{c|}{\textbf{child}}  \\ \hline
Female & 5.05 & 24.38 & 31.68  \\ \hline
Male & 2.82 & 19.45 & 30.10  \\ \hline
\end{tabular}
%}
\label{tab:EER}
\end{table} 

For the corpus collection, the disguise strategy was given to the speakers as an easy-to-understand and easy-to-execute type of speech modification, enabling a common condition to study the vocal variations. However, in our data collection process we did not consider how \emph{believable} the speakers are at producing disguised voices. Therefore, the present study aims to investigate how convincing the disguised samples sound and to evaluate the speaker's ability to disguise their voice by means of age estimations performed by a listener panel.

The perceptual experiment designed for the study aims to answer the following questions:
\begin{itemize}
 \item How accurate are human listeners at chronological age estimation from \textbf{modal} and \textbf{acted} voices?
 \item How successful are the speakers in modifying their voices in the 'intended' direction, in terms of perceived age by the listener panel?

\end{itemize}

The goal to evaluate the effectiveness of voice disguise is two-fold. On the one hand, it can give a perspective of how high is the threat of disguised speech to speaker recognition systems and how likely speakers are able to evade recognition with malicious intentions.  On the other hand, if the user needs to hide his identity for legitimate reasons, effective disguise could help in protecting the speaker's privacy and identity.

\section{Perceptual age estimation from speech}

Humans infer speaker characteristics from the voice on daily non face-to-face interactions. For instance, listening to speech relayed through a public address system, radio program, or speech interface gives the listener an impression of the speaker's gender, age, language, dialect, level of education, personality, among other factors. We focus on listener's ability to predict speaker's age from voice. 

Age prediction from one's voice has been addressed in a number of previous studies. Goy \emph{et.al.} \cite{Goy2016} studied age-related differences between older and younger speakers, and also listeners in terms of perceived speech and voice quality. In their experiments, the listeners estimated age and gender of young and old speakers, along with naturalness, clarity and intelligibility. The perception of voice quality was found to be significantly influenced by the age of the listeners. Vowel samples were used for the age estimation experiments and it was found that their younger listeners were more accurate.

Pettorino and Giannini \cite{pettorino2011speaker} addressed the degree to which listeners are able to effectively estimate the speakers' age. One of their experiments found that estimating the speaker's age in an unconstrained manner is a difficult task, while classifying a voice directly by age groups was relatively easy.

Age estimation from speech is of interest not only in defining high-level speaker characteristics but in the understanding of the changes related to aging. The aging process is not uniform and several extrinsic factors may affect speaker's voice as he or she ages. The work by Sch\"otz \cite{Schötz2007} presents an acoustic-phonetic study of the speaker's age. The study examined acoustic parameters of the voice such as speech rate, sound pressure level (SPL) and fundamental frequency ($F0$). These have been found important as acoustic correlates of the speaker's age. However, there is no clear relation between perceptual cues and listeners strategies used in age estimation and the age-related acoustic correlates. Also, other factors related to the speech sample, listening condition and listener's age have been found to have an effect on the human perception of age \cite{Schötz2007,Skoog2016}.   In age estimation by listening, it has been found that the age of young speakers tends to be overestimated, while the age of older speakers tend to be underestimated \cite{Skoog2016,Skoog2015can}.

Previous work has also addressed the effect of age disguise in the estimation of the speaker's age \cite{Skoog2016,Lass1982}. Skoog and Eriksson \cite{Skoog2016} studied the voice disguise of speakers that attempt to sound 20 years older and 20 years younger. It was found that the listeners' perceived an age change of 3 years, rather than the expected 20 years. In contrast, our study aims to evaluate whether the voice disguise attempts are perceived in the direction of the target age group, i.e. if the speakers were successful in the intended age modifications.

\section{Perceptual experiment}
We designed a perceptual test to study the quality of disguised voices using human listeners. The goal was to evaluate the disguise attempts through age estimation based on the speaker's voice. To this end, we first need three different definitions of \emph{age}:
\begin{mdframed}[style=MyFrame]\label{def:age-definitions}
\begin{center}
\begin{description}
	\item \textbf{Chronological age:} Objective age defined as the person's age at the time of the speech recording. We define this in years.
    \item \textbf{Perceived chronological age:} Subjective age rating by one listener concerning a given speech segment that reflects the listener's best guess of the actual chronological age. Differently from the actual chronological age, we define this age in terms of age categories.
    \item \textbf{Perceived intended age:} Categorical subjective age rating similar to the previous, except for one key difference: it is the listener's best guess of what age the speaker has \emph{intended} to sound like. Such variable can be meaningfully defined only for listeners who are aware of the presence of voice acting.
\end{description}
\end{center}
\end{mdframed}

The listeners chose their estimations of the speaker's perceived chronological age and perceived intended age from five pre-defined age intervals:  \emph{child} (younger than 18 years old), \emph{young adult} (approx. 18-30 years old), \emph{middle-age} (approx. 31-64 years old), \emph{retired} (approx. 65-80 years old) and \emph{senior citizen} (older than 81 years old). These intervals were determined empirically to have a balanced division of the speakers' chronological ages. The two boundary choices, younger than 18 years (child) and older than 80 years (elderly), are included so that for any speaker in our data, the listener has a chance to make a `correct' age estimation in the case of \emph{successful} age modification. For example, for a younger speaker that is able to modify the voice to sound younger, a listener could assign the child category.

\begin{figure}[!ht]
\centering
\includegraphics[width=0.6\columnwidth]{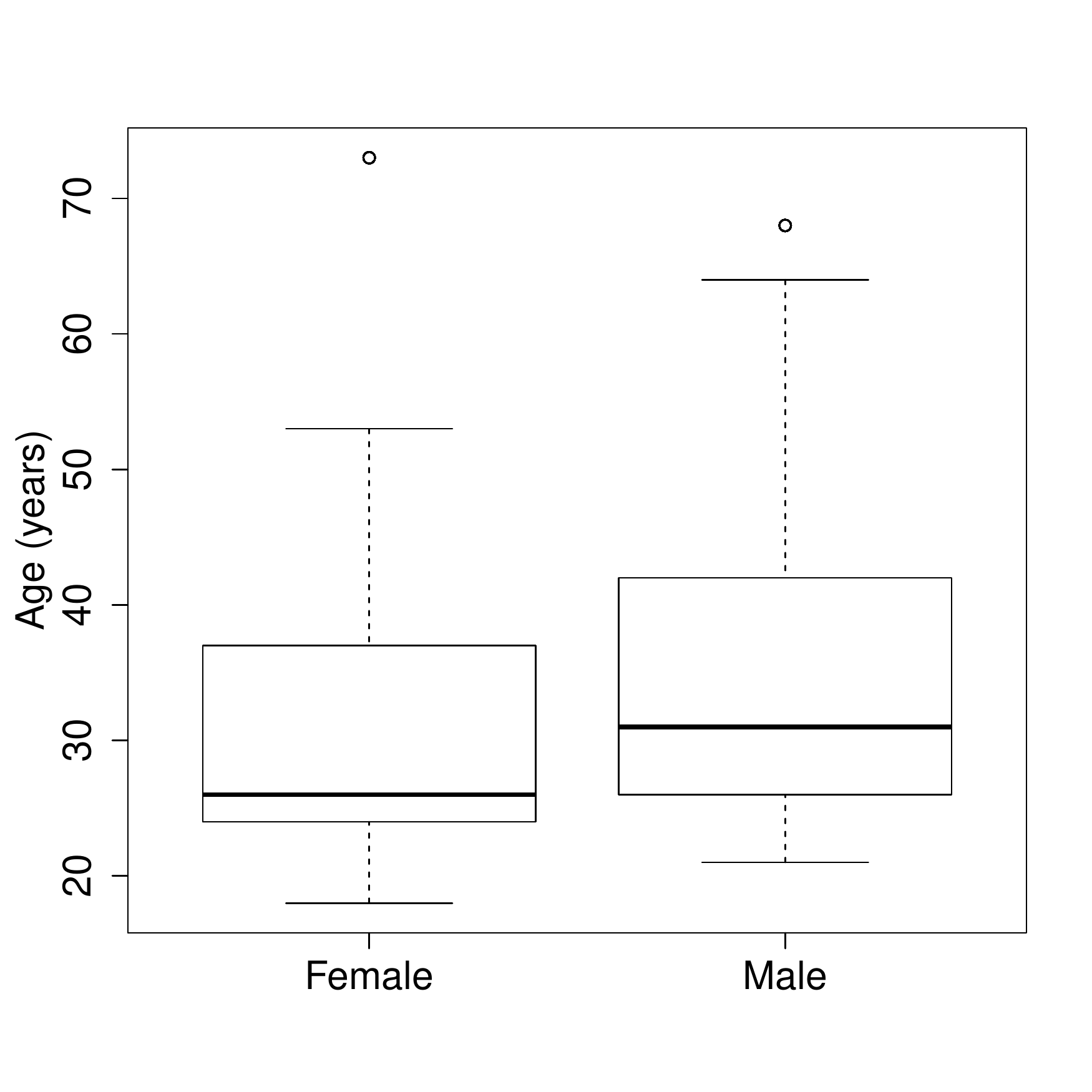}
\caption{{\it Speakers' chronological age distribution per gender (31 female and 29 male).}}
\label{fig:ages}
\end{figure}

Table \ref{tab:spkCategories} shows the speakers' distribution in the pre-defined age intervals and Figure \ref{fig:ages} shows the distribution of the speakers' chronological ages. The speech data for the listening test was recorded with a close-talking microphone and from the second recording session, where the speakers were generally more comfortable with the tasks. The data is the same as in our recent prior work \cite{hautamakiage,gonzalezhautamaki2017}; more details of the data collection can be found therein.  

\begin{table}[!htp]
\caption{{\it Speakers' age distribution according to the age categories used for the perceptual test.}}
\centering
%\resizebox{\linewidth}{!}{
\begin{tabular}{|l|c|c|}
\cline{2-3}
\multicolumn{ 1}{l}{}& \multicolumn{ 1}{|c|}{\textbf{Female}} & \multicolumn{1}{c|}{\textbf{Male}} \\ \hline
Younger than 18 years & 0& 0  \\ \hline
18 - 30 years & 19 & 14  \\ \hline
31 - 64 years & 11 & 14  \\ \hline
65 - 80 years & 1 & 1   \\ \hline
Older than 81 years & 0& 0  \\ \hline
\end{tabular}
%}
\label{tab:spkCategories}
\end{table} 

A set of 540 speech segments were selected for the perceptual test corresponding to equal number of segments from 60 native Finnish speakers in their modal, elderly and child voices. Three utterances were selected from each voice type per speaker. Therefore, $60$ speakers $\times$ $3$ voice types $\times$ $3$ utterances = $540$ utterances in total. The speech segments from all 60 are distributed in 12 speaker-disjoint folds, each consisting of 45 speech segments, from different speakers, with the same number of segments in modal and disguised voices ($15$ utterances $\times$ $3$ voice types = $45$). The 12 folds are then presented, likewise, to 12 independent listener panels. The listeners were informed that \emph{all} the speech segments contain voice acting --- in this way, the listeners will focus on the perceived age estimation regardless of whether a particular sample is acted or not.  

All the listener responses were collected with the aid of crowdsourcing. We had two groups of crowdworkers: those recruited ourselves, and those recruited through a crowdsourcing service. Concerning the recruited listeners by the authors, we prepared an online survey form that was completed by a total of 22 listeners. Each participant assessed 45 speech samples.
Even though such listeners participation is not considered a laboratory test, the listeners have participated in our previous voice comparisons tests, hence we regard them as reliable collaborators.

The advantage of using a paid crowdsourcing service is to reach a larger number and more diverse pool of participants in a short time.
To this end, 69 listeners participated in the listening test via Amazon mechanical turk\footnote{https://www.mturk.com/} (AMT) service. From these listeners, 35 listened to all the 45 samples (the same number as the recruited listeners), 23 listeners assessed more than 10 samples, and the remaining listeners less than 8 samples. All the listeners from the AMT group are non-native Finnish speakers while 11 listeners from the recruited group are native Finnish speakers.

\section{Results}

Listeners estimated speaker's age-group (five in total) for each speech sample. However, we are more interested in the perceived age in years rather than the number of votes per age group. We resort to estimate the {\em expected perceived age} per speaker, using the number of votes per age group as a weight. Let $x_i$ be the center of mass of age group $i \in \{1,2, \ldots , N\}$, $v_i$ number of votes in age group $i$ and $V$ the number of all votes for this speaker. In our case, the total number of age groups is $N=5$. The expected perceived age $a$ for a single speaker is then defined as:
\\
\begin{equation}
a = \frac{1}{V}\sum_{i=1}^N x_i v_i .
\label{eq:page}
\end{equation}
\\
The rest of this study will consider this as the age estimate given by the listeners collectively.

\subsection{Listener accuracy}
\begin{figure}[ht]
\includegraphics[width=\columnwidth]{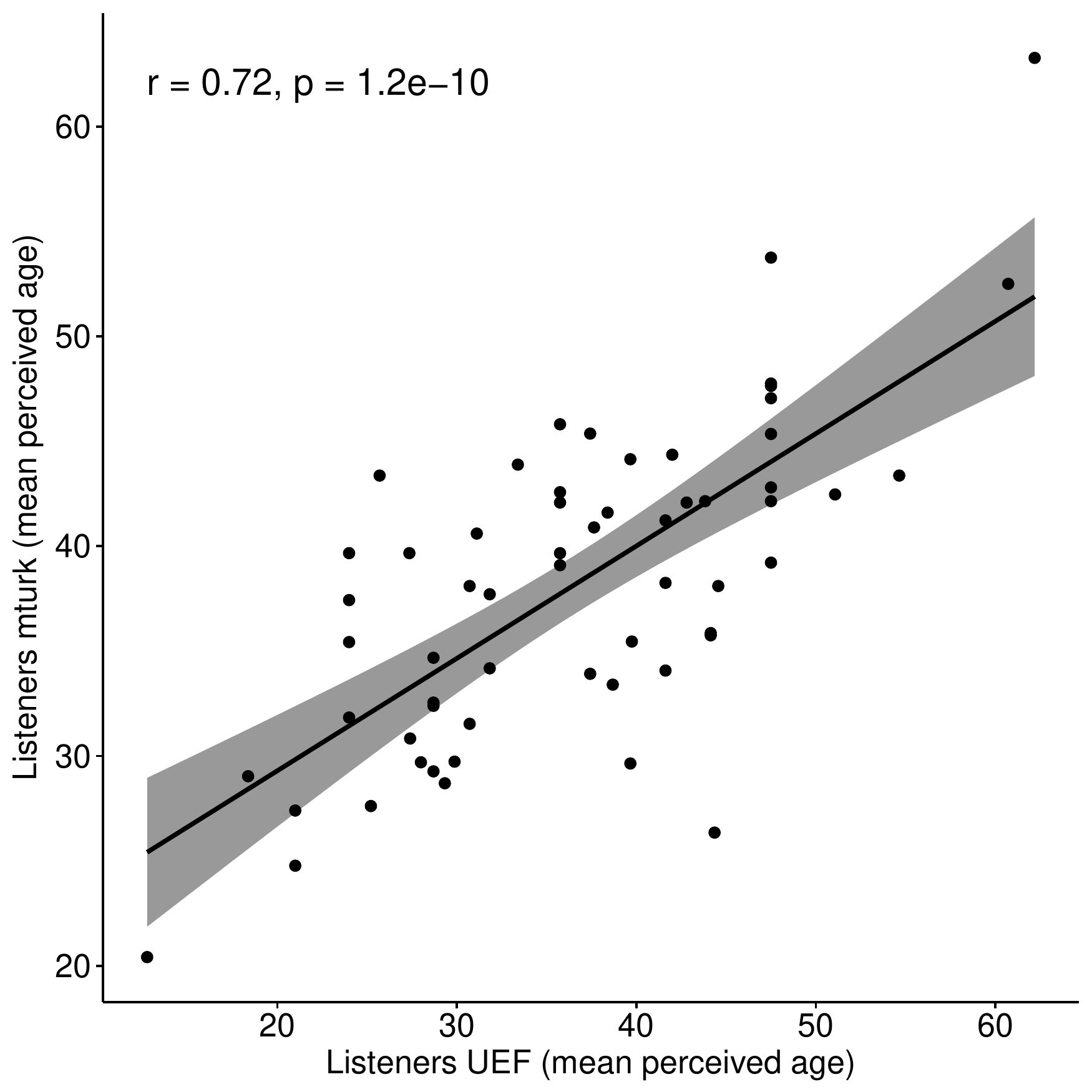}
\caption{{\it Pearson correlation of the speakers' perceived chronological age for the modal voice by the recruited listeners (UEF) and listeners from the crowd-source platform (Amazon mechanical turk). Correlation coefficient $0.72$, {p-value} $1.2e^{-10}$ with a confidence interval of $[0.5658~,~0.8212]$ at 95\%.}}
\label{corr_list}
\end{figure}

We evaluated the estimations obtained by the listeners recruited by the authors (UEF) and the ones from crowdsourcing (AMT). This comparison provides information of the similarity of the responses by both groups of listeners. Using the modal voice samples, the mean perceived chronological age for each speaker was obtained using Eq. (\ref{eq:page}), then the correlation between the two listeners groups estimations was calculated.  Figure \ref{corr_list} was generated using \texttt{ggpubr} R package. It presents the correlation using \emph{Pearson} method where UEF listeners and AMT listeners estimations are significantly correlated with a coefficient of $0.72$ and \emph{p-value} of $1.2e^{-10}$. Indicating a positive correlation between the variables. The gray area shows the uncertainty of the correlation coefficient at the 95\% confidence interval.

Even though the listener groups are of different sizes, the perceived age estimations themselves are comparable. We can therefore expect similar performance between random listeners from the UEF and the AMT groups. For the rest of the analysis, therefore, we pool the recruited and the AMT listeners into one listener panel of 91 listeners.

To evaluate the age estimations for each speech sample, the \emph{age difference} was calculated as follows:\\

\textit{Age diff.} $=$ \textit{Chronological age} $-$ \textit{Perceived age}, \\

\noindent where \emph{perceived age} corresponded to either the perceived chronological age or to the perceived intended age.
A positive age difference can be interpreted as the perceived age being underestimated or lower than the speaker's chronological age, and a negative value would correspond to an overestimated perceived age, or as higher than the speaker's age.
%Probably too late to change now but to me perceived minus chrono would be more intuitive: in that case values > 0 indicate "overestimation" and < 0 "underestimation", with 0 being correct estimation.

To evaluate listeners ability to estimate speaker's chronological age, we compared perceived chronological and intended ages to the real age of the speaker only for the modal voice speech segments. The assumption is that for each speaker the perceived chronological age and the perceived intended age will be similar or show a small difference, as these samples do not include disguised voices.

\begin{figure}[t!]
\begin{minipage}[b]{1.0\linewidth}
  \centering
   \centerline{\includegraphics[width=0.9\linewidth]{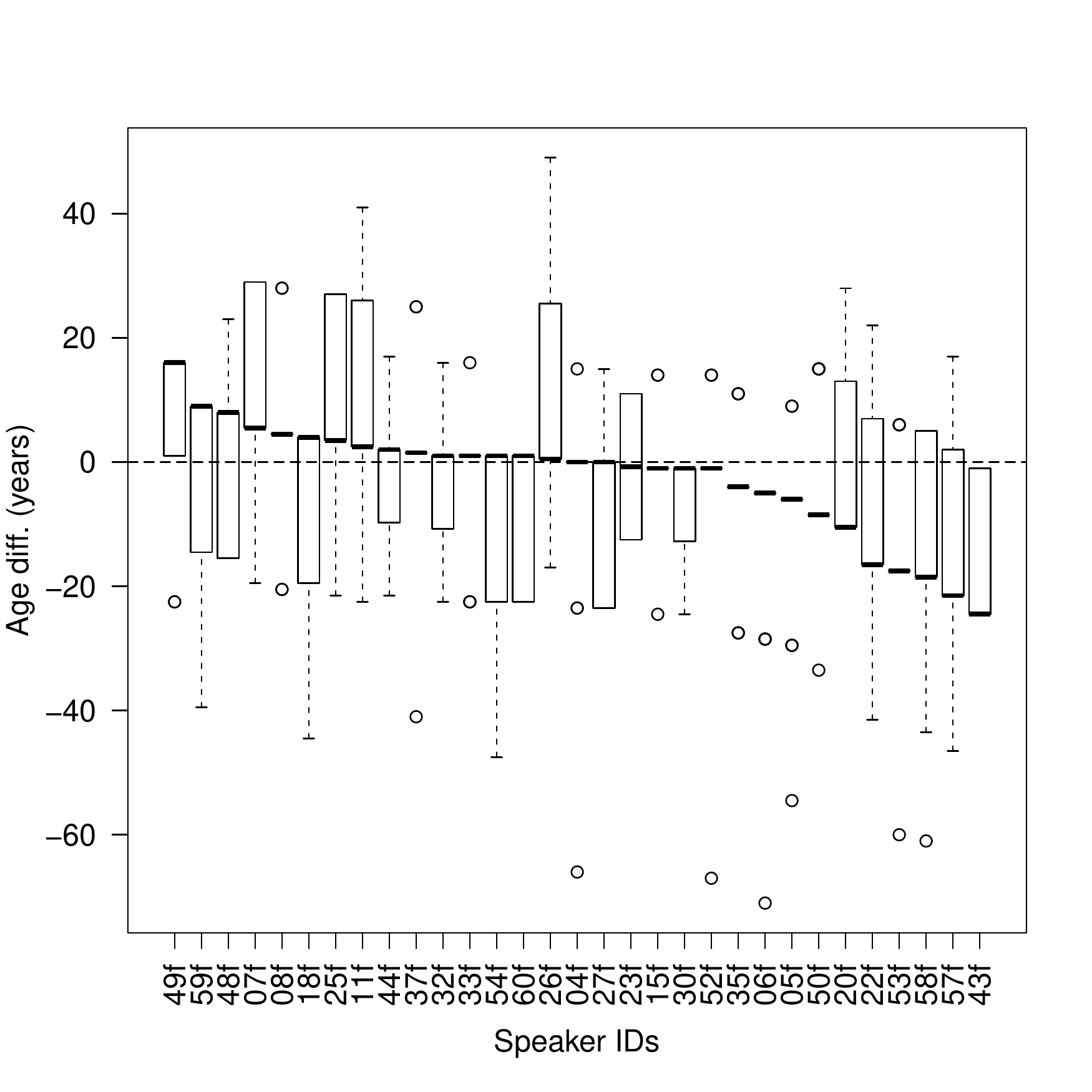}}
  \centerline{{\it(a) Difference for chronological age estimates}}\medskip
\end{minipage}

\begin{minipage}[b]{1.0\linewidth}
  \centering
  \centerline{\includegraphics[width=0.9\linewidth]{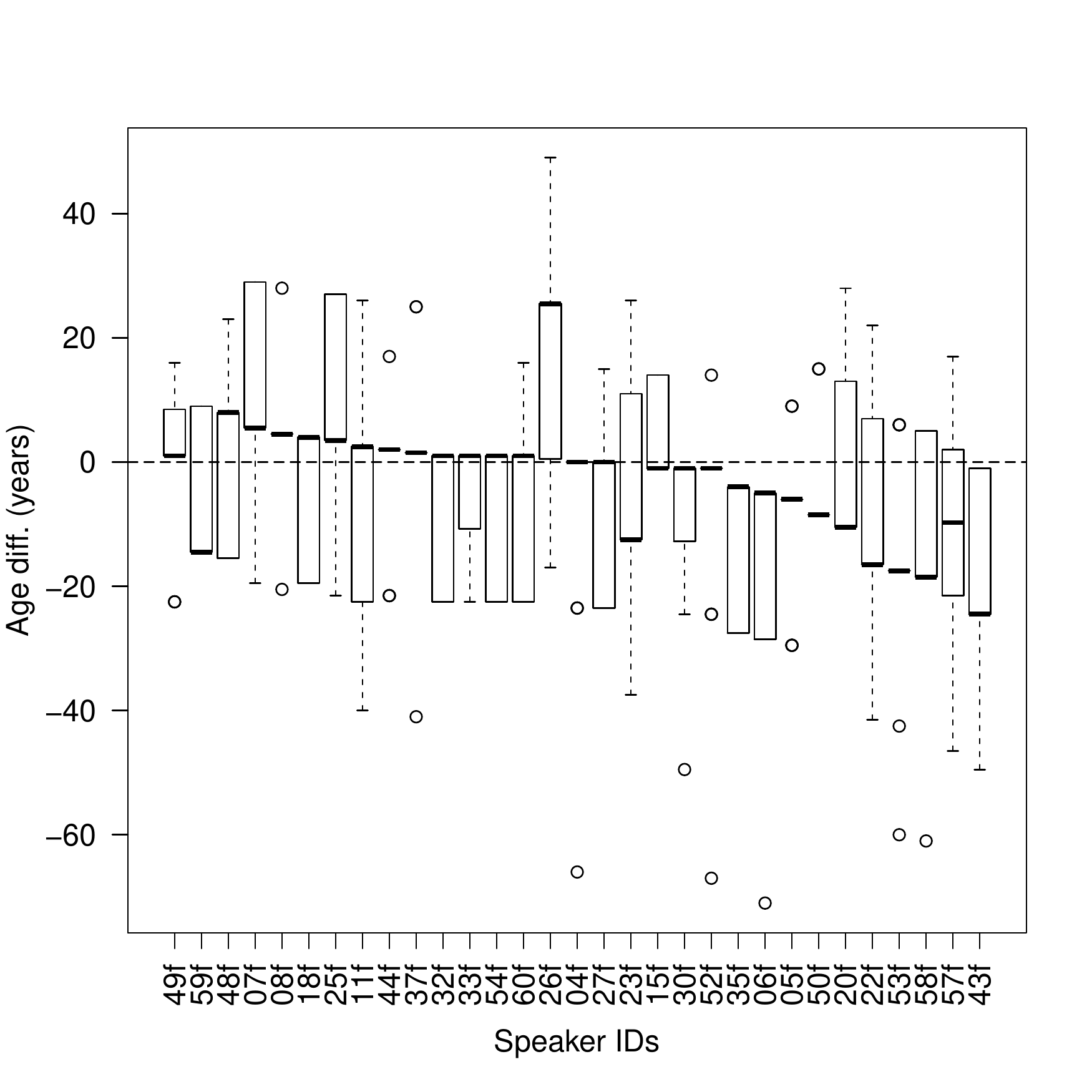}}

  \centerline{{\it(b) Difference for perceived intended age}}\medskip
\end{minipage}
\caption{{\it Female speakers age difference in years between the speakers' chronological age and the perceived chronological and perceived intended age estimates for the modal voice segments. The speakers are ordered by the median age difference in descending order. The graphs show small differences indicating that the estimates for the age of the speaker and the ``intended'' age are close for modal voice.}}
\label{fig:Fmodal}
\end{figure}

Figure~\ref{fig:Fmodal} shows the perceived age estimations for female speakers in the case of modal voices. The median age differences are close to the zero difference region for many speakers. This result indicates that the listeners' perceived age estimations are close to the speakers' chronological ages and that the perceived chronological and intended age estimations are similar for the modal voices. This agrees with our assumption for most of the speakers.  
\begin{figure}[t!]
\begin{minipage}[b]{1.0\linewidth}
  \centering
   \centerline{\includegraphics[width=0.9\linewidth]{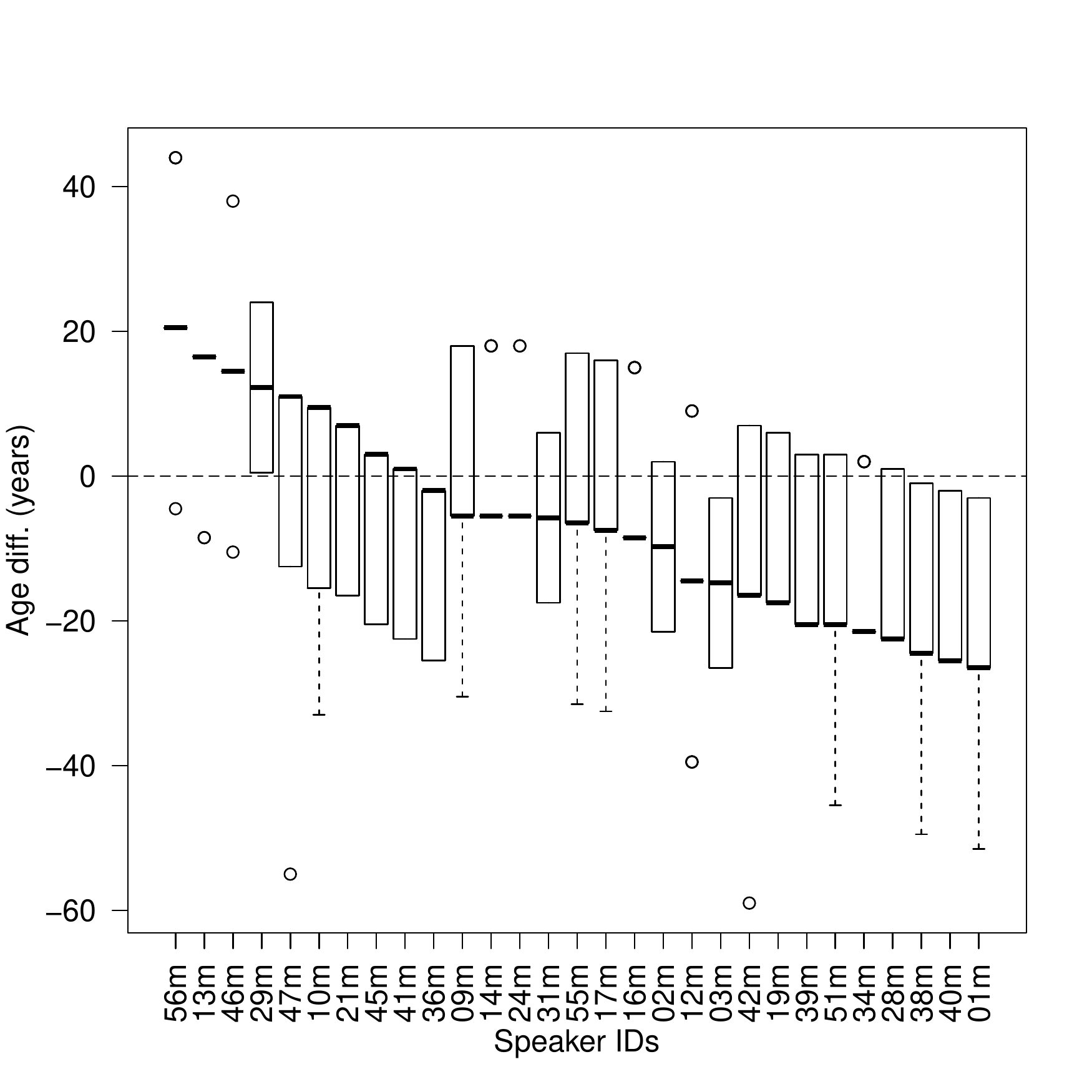}}
  \centerline{{\it(a) Difference for chronological age estimates}}\medskip
\end{minipage}

\begin{minipage}[b]{1.0\linewidth}
  \centering
  \centerline{\includegraphics[width=0.9\linewidth]{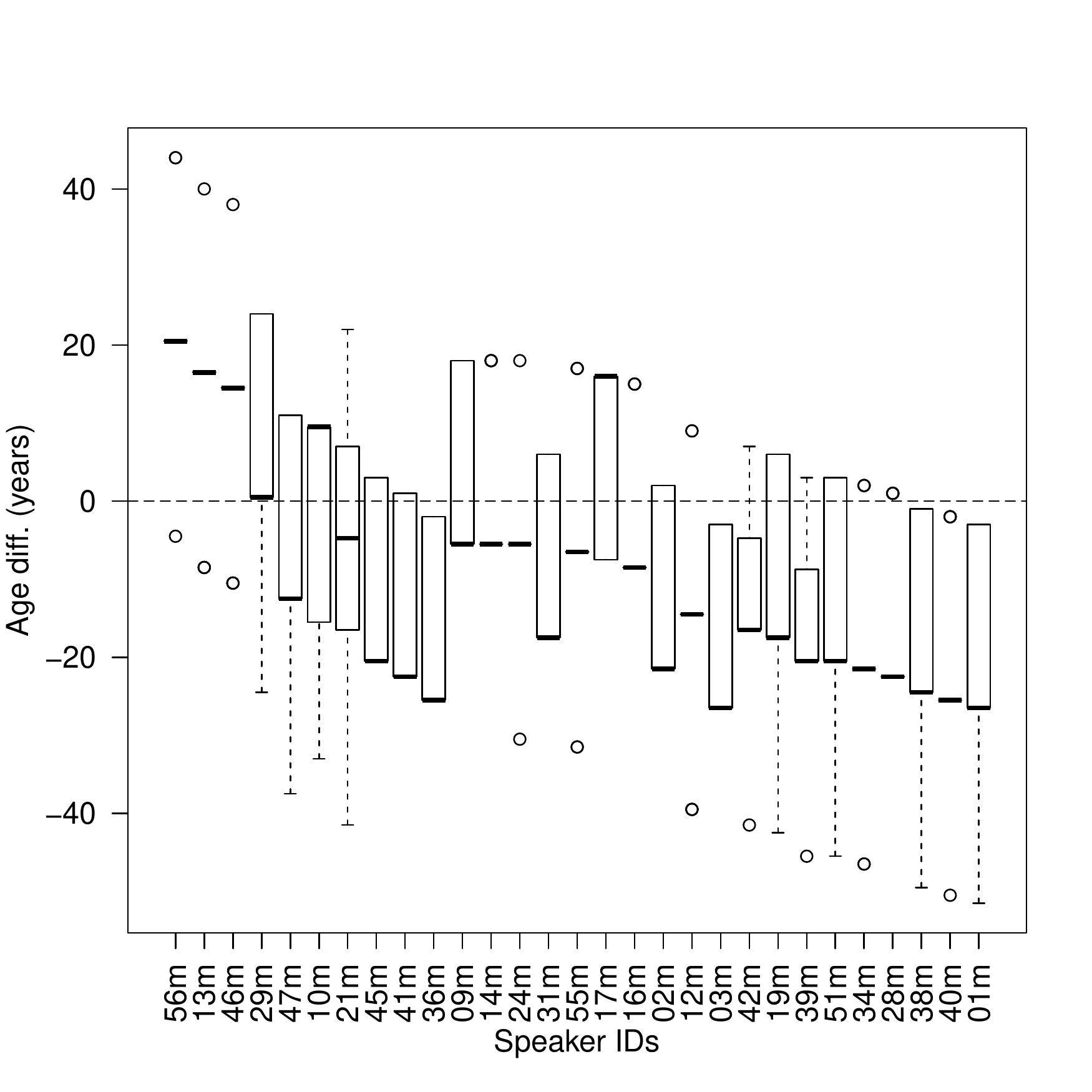}}

  \centerline{{\it(b) Difference for perceived intended age}}\medskip
\end{minipage}
\caption{{\it Male speakers age difference in years between the speakers' chronological age and the perceived chronological and perceived intended age estimates for the modal voice segments. The speakers are ordered by the median age difference in descending order. The graphs show a similar pattern in the age differences, in both graphs the estimations are mostly to the negative side indicating that the age of the speakers is over-estimated.}}
\label{fig:Mmodal}
\end{figure}

The results for male speakers are shown in Figure~\ref{fig:Mmodal} where the differences for perceived chronological (Fig.~\ref{fig:Mmodal} (a))  and intended age (Fig.~\ref{fig:Mmodal} (b)) follow the same pattern but, in contrast to female speakers, the difference is not close to the zero difference region and it is negative for most of the speakers.  This indicates that for most of the speakers, the listeners over-estimated the ages, which is a common problem on age estimation of young speakers. 

Age estimation from speech is a difficult task and our experiment has the additional challenge of including voice disguise by age modification. For our corpus, our listeners were better at estimating the chronological age of female speakers than male speakers. Even though the age estimation show variations for each speaker, we can consider the performance of the listeners to be consistent in the estimation of chronological age from modal and disguised voices for most of the speakers.

%Chronological Age vs. perceived acted age for elderly voice.\\
\begin{figure}[!t]
\begin{minipage}[b]{1.0\linewidth}
  \centering
   \centerline{\includegraphics[width=\columnwidth]{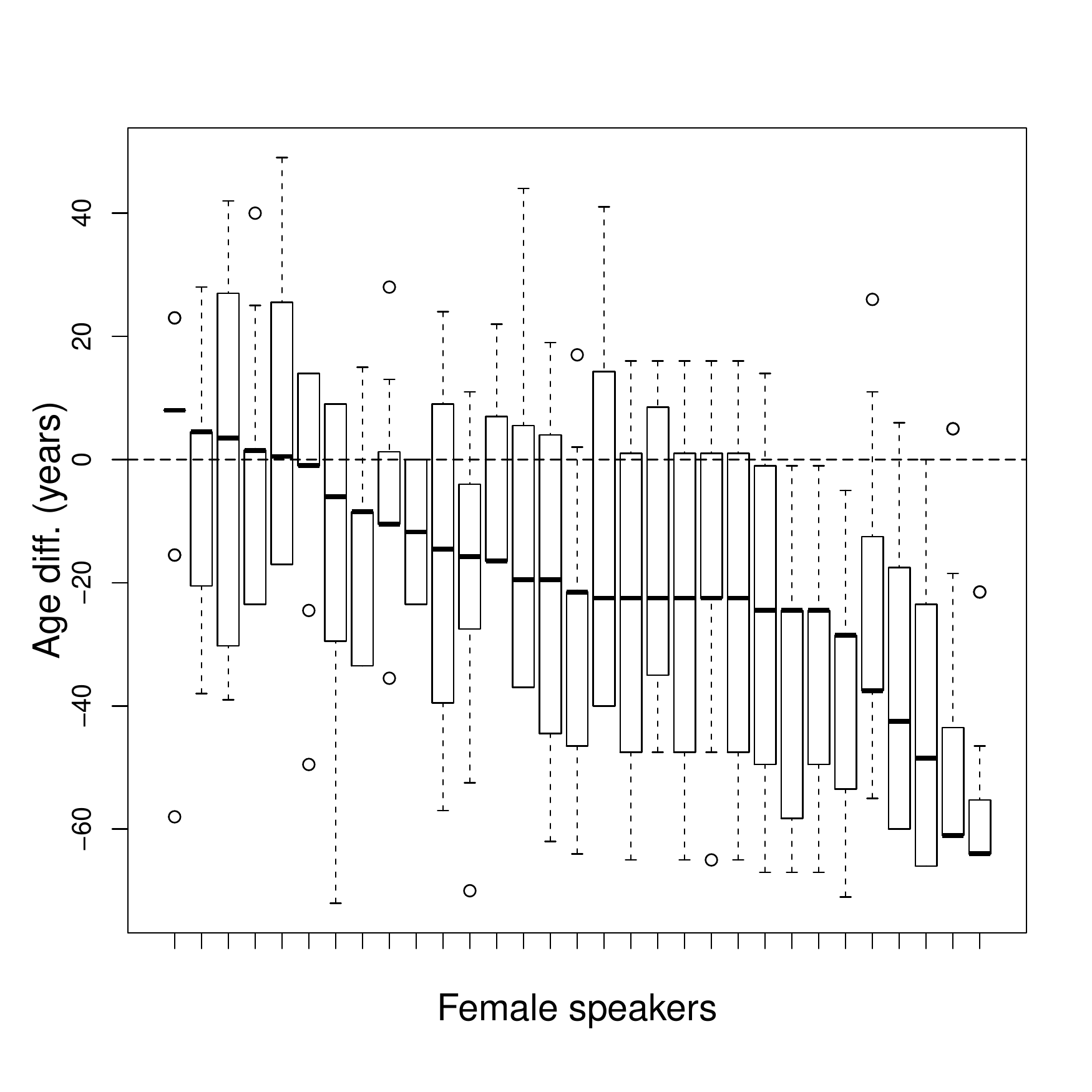}}
  %\centerline{(a) Female }\medskip
\end{minipage}

\begin{minipage}[b]{1.0\linewidth}
  \centering
  \centerline{\includegraphics[width=\columnwidth]{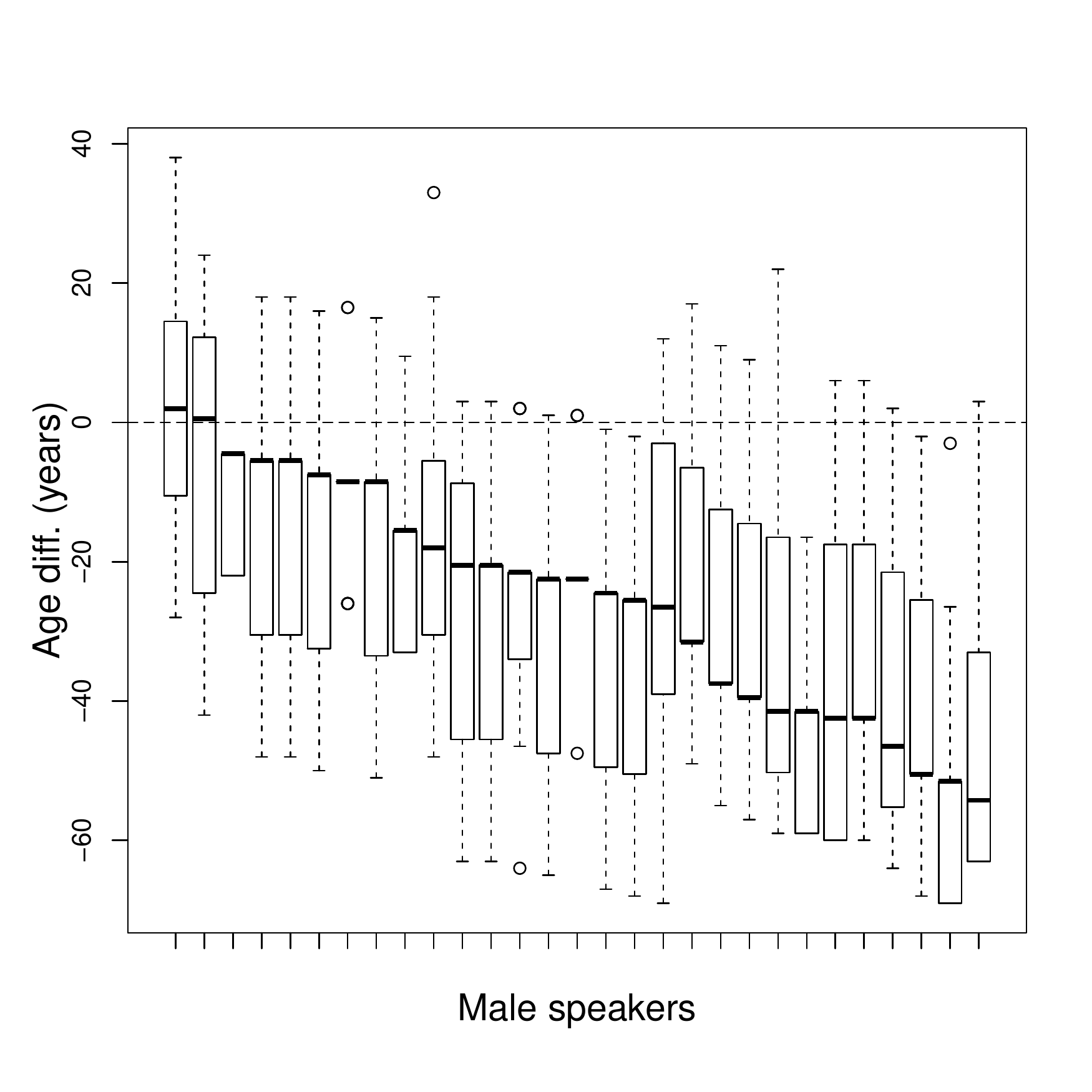}}

 % \centerline{(b) Male }\medskip
\end{minipage}
\caption{{\it Age difference between the speakers' chronological age and perceived intended age for the intended \emph{elderly} voice segments per gender. The speakers are ordered by the median age difference in a descending order.}}
\label{fig:Oldvoice}
\end{figure}

%Chronological Age vs. perceived acted age for child voice.\\
\begin{figure}[th!]
\begin{minipage}[b]{1.0\linewidth}
  \centering
   \centerline{\includegraphics[width=\columnwidth]{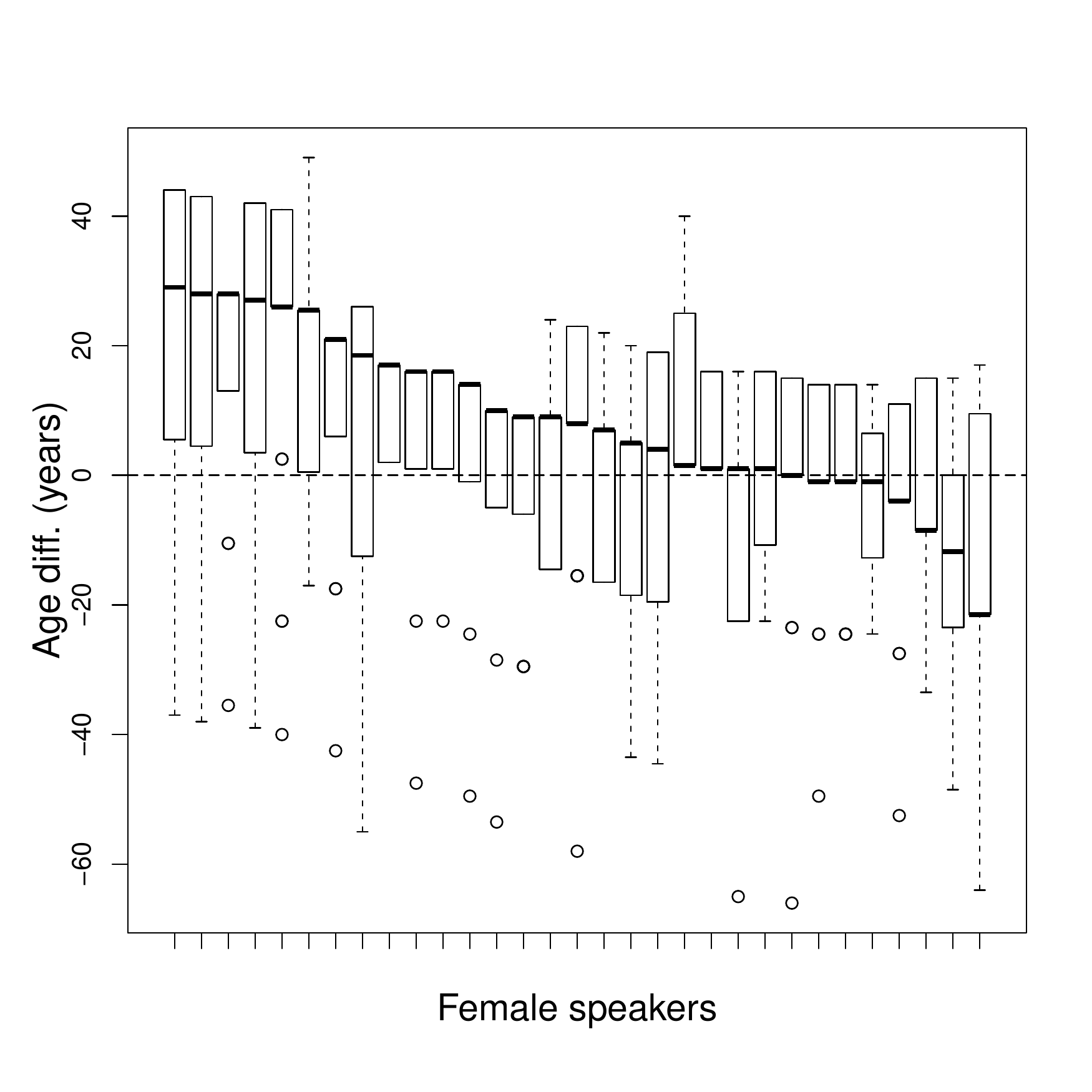}}
  %\centerline{(a) Female }\medskip
\end{minipage}

\begin{minipage}[b]{1.0\linewidth}
  \centering
  \centerline{\includegraphics[width=\columnwidth]{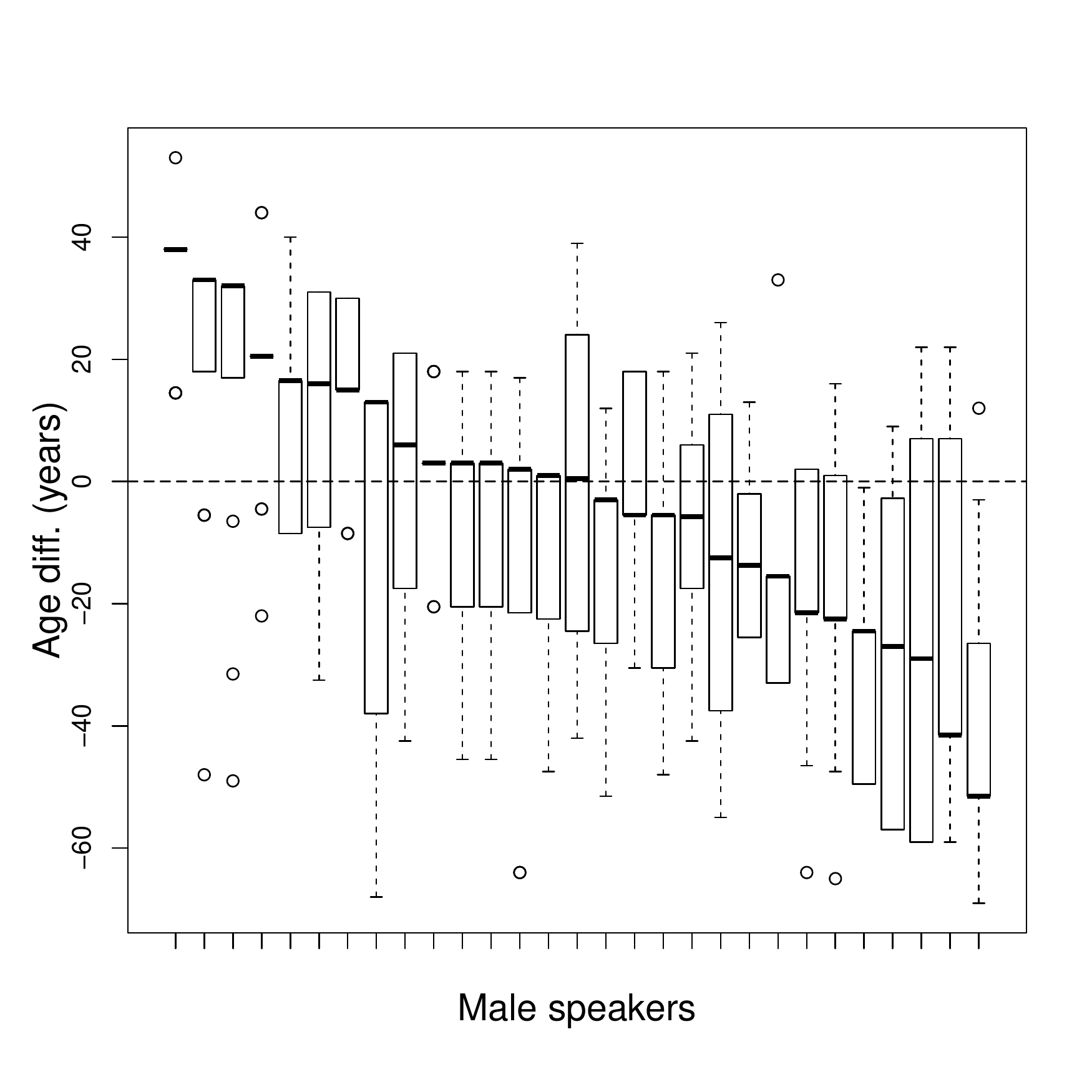}}

  %\centerline{(b) Male }\medskip
\end{minipage}
\caption{{\it Age difference between the speakers' chronological age and perceived intended age for the intended \emph{child} voice segments. The speakers are ordered by the median age difference in a descending order.}}
\label{fig:Youngvoice}
\end{figure}

\subsection{Perceived age of disguised voices}
Similar to Fig. \ref{fig:Fmodal} (b), let us now consider the samples containing disguised voices. The assessments were carried out using the age difference between the chronological age of the speakers and the perceived intended age. In this way, the listeners' age estimations will reveal whether the speakers' attempts reached the intended voice. 

%FemaleOldActed --> age diff mean range[2.46 -57.12] MoM=-19.45
%MaleOldActed--> [1.36 -46.53] MofM = -24.97
Figure~\ref{fig:Oldvoice} shows the age difference in years for the speech samples with intended elderly voice. For female speakers, the mean age difference range is from 2.46 to -57.12 with a mean of -19.45, while for male speakers, the range is from 1.36 to 46.53 with a mean of -24.97.  We observe a negative difference for both genders, suggesting that the listeners estimated the perceived intended age to be older than the speaker's chronological age. This can be considered as a successful age disguise attempt in terms of the perceived age. Even though most of the speakers were able to sound older than themselves, the age difference is small to reach the intended age (elderly or older than 80 years old). 

%FemaleYoungActed --> [20.81 -11.29] MoM= 5.08
%MaleYoungActed--> [33.61 -38.93] mean of means = -5.95
For the intended child voice, the age difference is expected to be positive to be considered successful age disguise: this means that the age of the perceived intended voice is younger than the speaker's chronological age. Figure \ref{fig:Youngvoice} shows the age difference per gender for intended child voice. For female speakers, the mean age difference range is from 20.81 to -11.29 with a mean of -5.08, while for male speakers, the range is from 33.61 to -38.93 with a mean of -5.95. For female speakers, most of the perceived intended age estimates are below the speaker's chronological age, but just a few were in an age difference that would correspond to the child's voice.  This was not the case for male speakers, where the perceived intended age was overestimated with respect to the speaker's chronological age. In other words the listeners' estimations indicate that the voices corresponded to older voices and not the intended child voices.

\subsection{Chronological age estimation from disguised voices}
\begin{figure}[!ht]
\begin{minipage}[b]{1.0\linewidth}
  \centering
   \centerline{\includegraphics[width=\columnwidth]{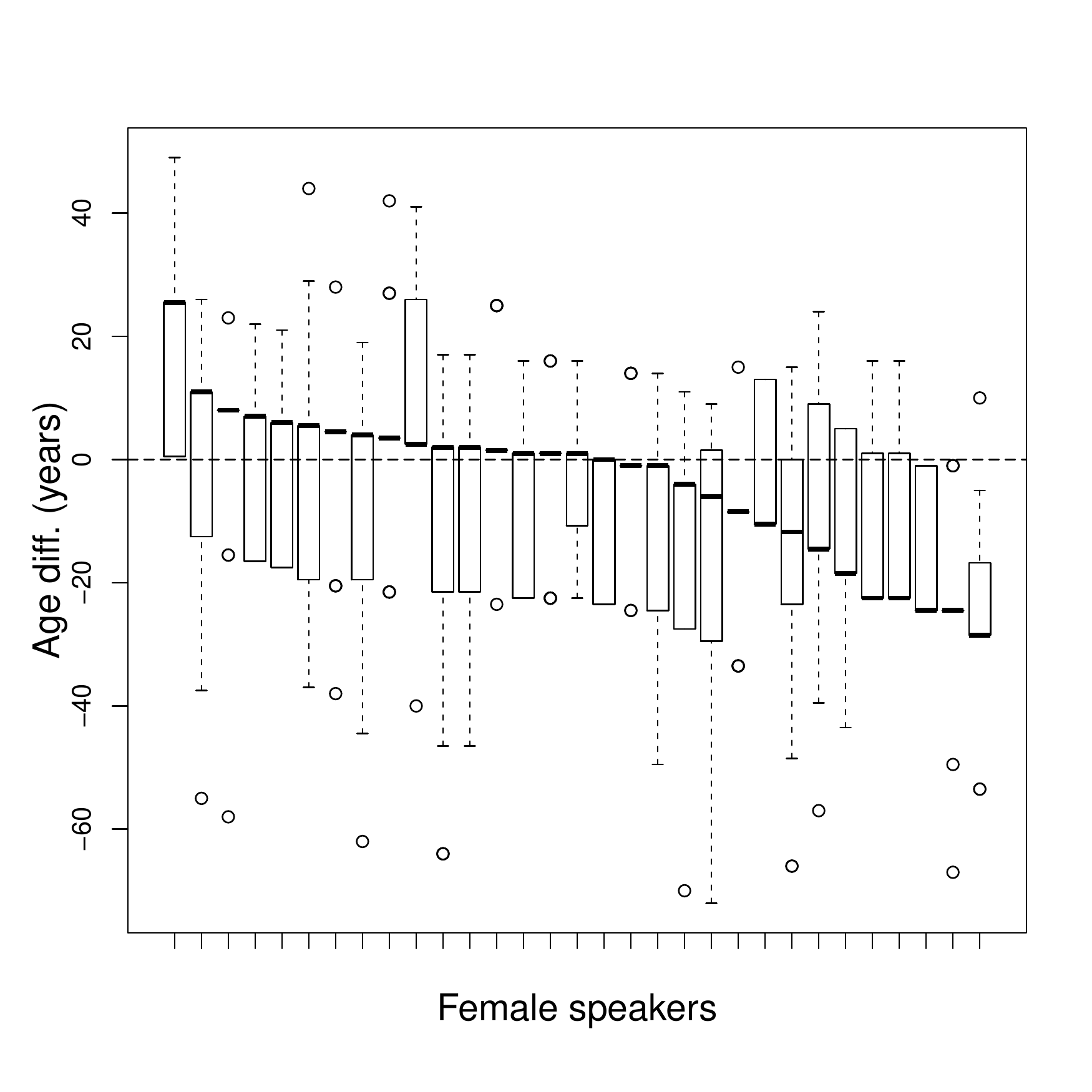}}
%  \centerline{(a) Female }\medskip
\end{minipage}

\begin{minipage}[b]{1.0\linewidth}
  \centering
  \centerline{\includegraphics[width=\columnwidth]{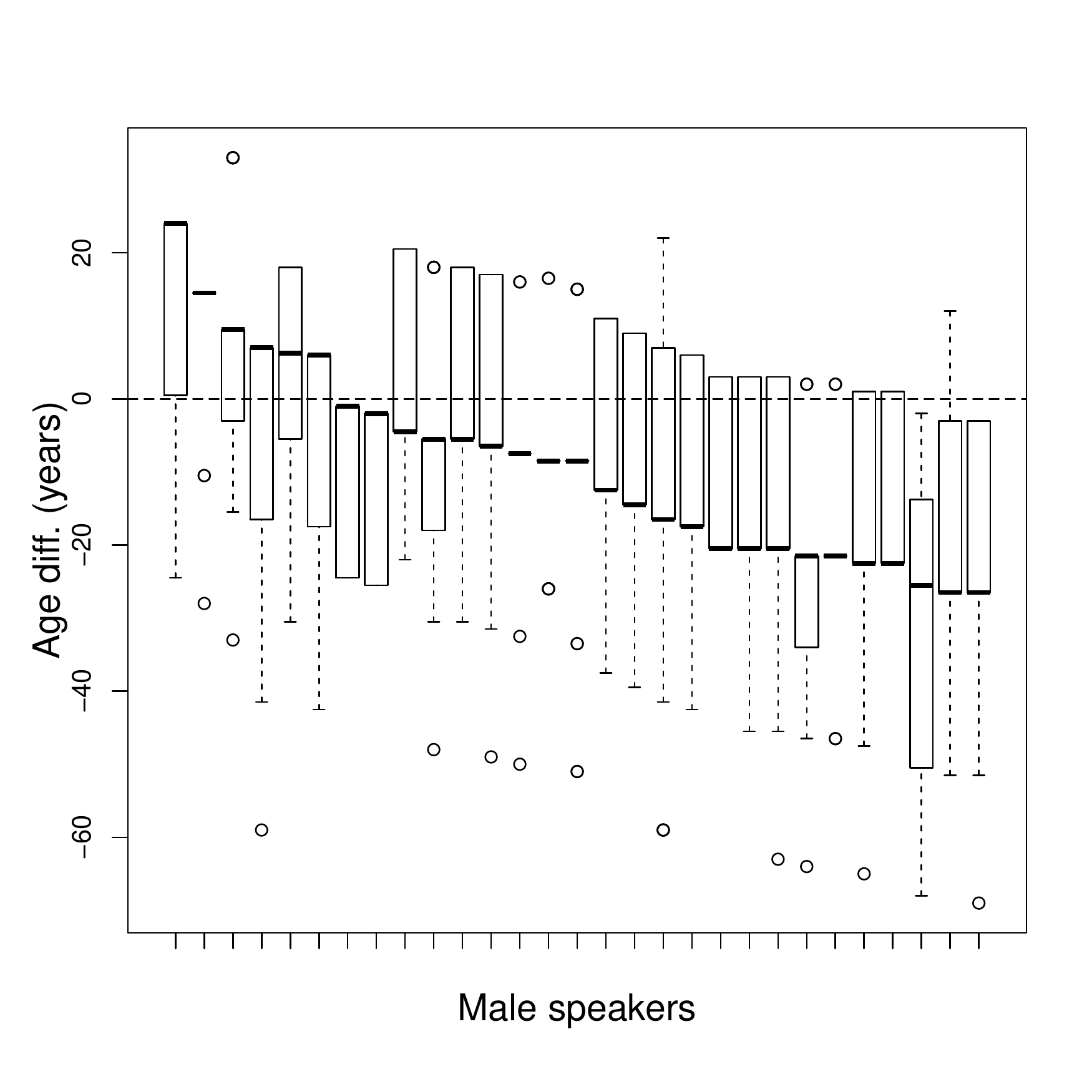}}

 % \centerline{(b) Male }\medskip
\end{minipage}
\caption{{\it Age difference between the speakers' chronological age and perceived chronological age for the intended \emph{elderly} voice segments. The speakers are ordered by the median age difference in a descending order.}}
\label{fig:Oldvoice_real}
\end{figure}

\begin{figure}[!ht]
\begin{minipage}[b]{1.0\linewidth}
  \centering
   \centerline{\includegraphics[width=\columnwidth]{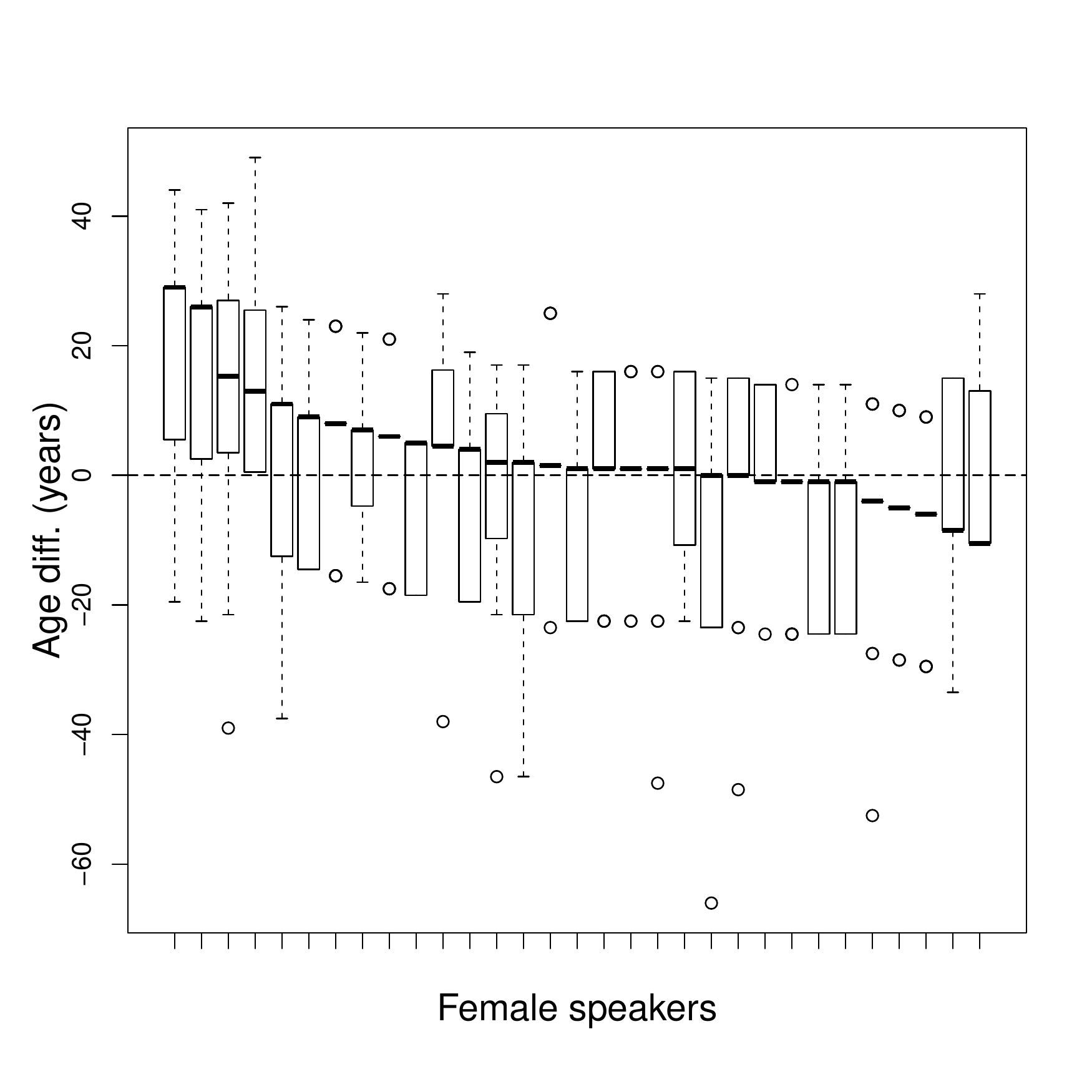}}
 % \centerline{(a) Female }\medskip
\end{minipage}

\begin{minipage}[b]{1.0\linewidth}
  \centering
  \centerline{\includegraphics[width=\columnwidth]{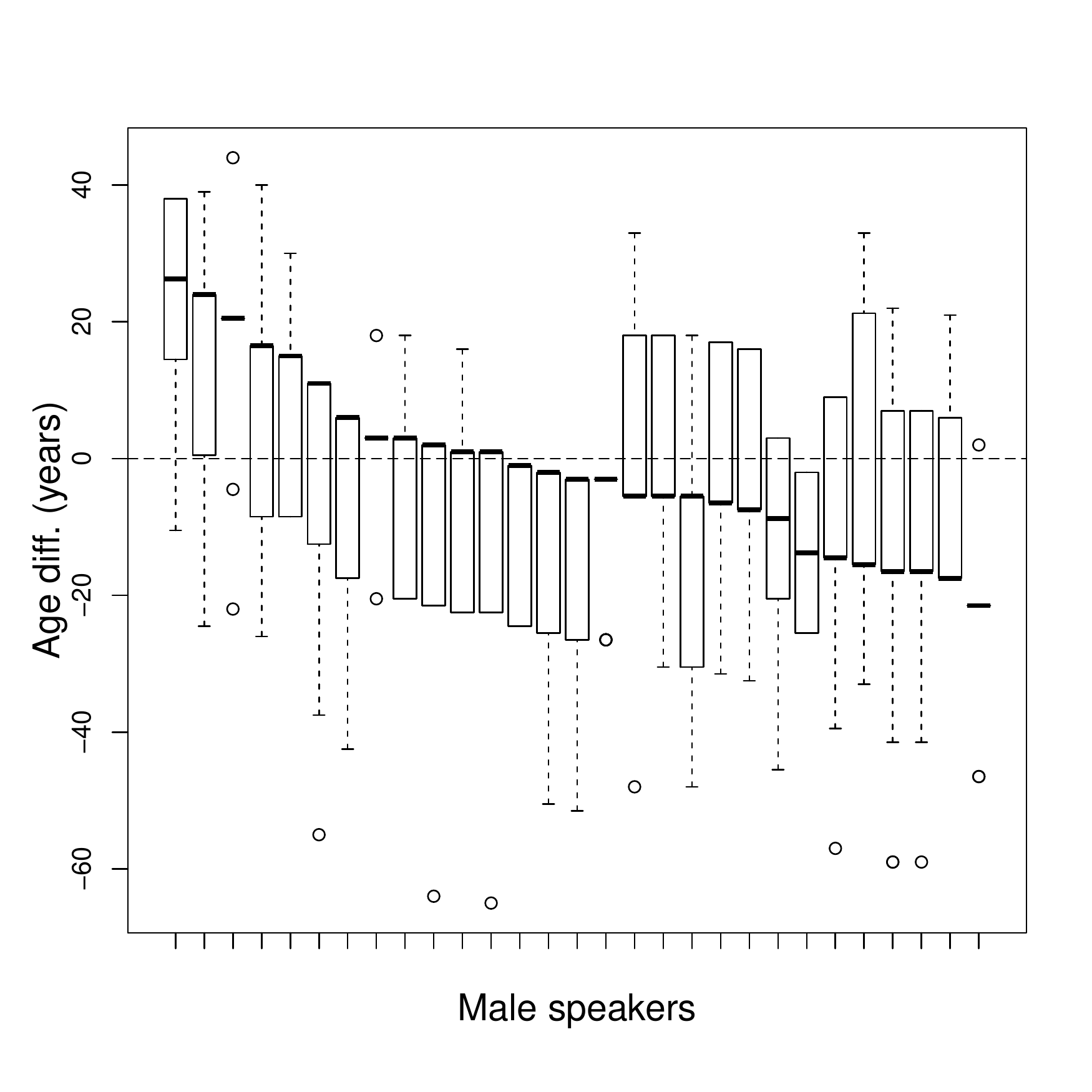}}

  %\centerline{(b) Male }\medskip
\end{minipage}
\caption{{\it Age difference between the speakers' chronological age and perceived chronological age for the intended \emph{child} voice segments. The speakers are ordered by the median age difference in a descending order.}}
\label{fig:Youngvoice_real}
\end{figure}

Another interesting point was to evaluate the estimation of the speaker's chronological age when he or she disguises the voice. This could give insights of how challenging the age estimation task is when the speaker is trying to disguise the voice by means of age modification. Our results, Figures \ref{fig:Oldvoice_real} and \ref{fig:Youngvoice_real}, show that for most of the male and female speakers, the perceived chronological age estimations were similar for both attempts of elderly and child voices. The results indicate that the listeners' age estimation only were affected for a few speakers but followed the results obtained from the perceived chronological age from modal voices.

%Chronological Age vs. perceived chronological age for child voice.\\

\section{Conclusions}
In this study, we looked into what a successful voice disguise attack on automatic speaker verification sounds like to humans. We approached this question by gathering human age estimations for modal and disguised speech samples and comparing these estimated ages against each other. Listeners estimated the chronological and the intended age of the speaker for the selected speakers' samples. This study also served as the authors' first perceptual experiment with a crowdsourcing platform, Amazon Mechanical Turk, to facilitate the convenient collection of listener results. We observed a positive correlation between the AMT responses to those collected locally with ad-hoc listener recruitment. This gives us confidence for the use of paid crowdsourcing services in future studies as well.

In our experiments, listeners were able to approximate the chronological age of female speakers with their modal voices, while for male speakers, the age was systematically overestimated. This is a common observation in previous studies for the age estimation of young speakers \cite{Skoog2015can}, which is the case of most of the speakers in this study. Also, the perceived chronological age from disguised voices followed similar estimations as with modal voice. 

In the evaluation of the disguised voices for female speakers, the listeners' estimations of the perceived intended age followed the direction of the target age for intended elderly and child voices. That was also the case of intended old voice for male speakers. However, the intended child voice was perceived as belonging to an old person for most of the male speakers. This indicates that even though intended child voice from most male speakers did not sound believable, this type of disguise affected the performance of ASV systems.

In the light of these results, future work could focus on the cues listeners detect from disguised speech to effectively identify that it is disguised. In general, listeners are able to distinguish between acted and non-acted voice, but a standard automatic speaker recognition system is not equipped to recognize speakers under these type of disguise strategies. 

\bibliographystyle{IEEEbib}
\bibliography{refs}

\end{document}